\documentclass{article}
\usepackage{ijcai19}

\usepackage{times}
\usepackage{soul}
\usepackage{url}
\usepackage[hidelinks]{hyperref}
\usepackage{url}  
\hypersetup{
    colorlinks=true,
    linkcolor=blue,
    filecolor=magenta,      
    urlcolor=magenta,
}
\usepackage[utf8]{inputenc}
\usepackage[small]{caption}
\usepackage{graphicx}
\usepackage{amsmath}
\usepackage{booktabs}
\title{KRNET: Image Denoising with Kernel Regulation Network}
\author{
Peng Liu$^1$\and
Xiaoxiao Zhou$^2$\and
Junyiyang Li$^{1}$\and
El Basha Mohammad D$^1$\and
Ruogu Fang$^1$\\
\affiliations
$^1$University of Florida, USA\\
$^2$Sun Yat-Sen University, China\\
\emails
\{pliu1, yangjunyili92,mdelbasha\}@ufl.edu,
zhouxx5@mail2.sysu.edu.cn,
ruogu.fang@bme.ufl.edu
}

\begin{document}

\maketitle

\begin{abstract}

One popular strategy for image denoising is to design a generalized regularization term that is capable of exploring the implicit prior underlying data observation.
Convolutional neural networks (CNN) have shown the powerful capability to learn image prior information through a stack of layers defined by a combination of kernels (filters) on the input. 
However, existing CNN-based methods mainly focus on synthetic gray-scale images. 
These methods still exhibit low performance when tackling multi-channel color image denoising. 
In this paper, we optimize CNN regularization capability by developing a kernel regulation module. In particular, we propose a kernel regulation network-block, referred to as KR-block, by integrating the merits of both large and small kernels, that can effectively estimate features in solving image denoising. We build a deep CNN-based denoiser, referred to as KRNET, via concatenating multiple KR-blocks. We evaluate KRNET on additive white Gaussian noise (AWGN), multi-channel (MC) noise, and realistic noise, where KRNET obtains significant performance gains over state-of-the-art methods across a wide spectrum of noise levels.
\textit{The code is available at \url{https://github.com/cswin/RC-Nets}}. 

\end{abstract}

\section{Introduction} \label{intro}
Digital images are playing an essential role in both daily life and industry more than ever before. Image denoising remains a fundamental study of image restoration and computer vision. It aims to recover the clean image $\mathrm{x}$ from a noisy observation \begin{equation}
    \mathrm{y=x+\eta}
\end{equation}
making an estimated clean image $\mathrm{ \widehat{x} }$ approaching $\mathrm{x}$ as much as possible, where $\mathrm{\eta}$ is the noise term; the object function is denoted as $\mathrm{\Phi(x;y)}$. In common, $\mathrm{\eta}$ is assumed  to be  additive Gaussian white noise (AGWN).

The contextual information from the data observation $\mathrm{(x;y)}$ is not sufficient to achieve a satisfying accuracy on the estimated $\mathrm{ \widehat{x} }$~\cite{geman1995nonlinear}. Therefore, a regularization term is usually needed to obtain a prior information of the data. The recovered $\mathrm{ \widehat{x} }$ is formulated as:
\begin{equation}
\label{eq:map2}
\begin{split}
 \mathrm{ \widehat{x}=arg\underset{x}{min}\Phi _{R}(x)+\lambda \Phi _{D}(x;y)}
\end{split}
\end{equation}
where $\mathrm{\Phi _{R}(x)}$ represents regularization terms such as the Wiener filter and Bayesian methods. $\mathrm{\Phi _{D}(x;y)}$ is fidelity  term to the data, and a common choice is likelihood, which is represented as $\mathrm{\Phi _{D}(x;y) =\left \| y-x \right \|^{2}}$; $\mathrm{\lambda}$ is a positive trade-off parameter to balance the weight between the two terms. Typically, to overcome the challenge of nonlinear estimators, auxiliary parameters $\Theta$ are introduced to have a new objective function as: 
\begin{equation}
\label{eq:map3}
    \mathrm{\Phi ^{*}(x,\Theta )=\Phi_{R}^{*}(x,\Theta)+\lambda \left \| y-x \right \|^2}
\end{equation}
More formally, from Bayesian perspective with the log-likelihood of Maximum A-Posteriori (MAP) Estimation: 
\begin{equation}
\label{eq:map4}
\mathrm{ \widehat{x}= arg\underset{x}{min}\frac{\lambda }{2}\left \| y-x\right \|^{2}+\Phi_{R}^{*}(x,\Theta)}
\end{equation}

One popular idea for overcoming the challenges of solving Eqn.~\ref{eq:map4} is to decouple the two terms in Eqn.~\ref{eq:map3} such that exploring priors  (regularization)  and enhancing model generality (likelihood) are independent. By combining with term decoupling,  one strategy is to train a deep convolutional neural network (CNN) as a prior integrating with model-based optimization methods (e.g.,~\cite{zhang2017learning}). Another strategy is to utilize recurrent network for exploring data priors and then combine with transfer learning~\cite{torrey2009transfer} for enhancing model generality (e.g.,~\cite{xiao2017discriminative}). However, the combination of two types methods usually introduces additional variables and computational cost for regulating the interaction between them.

CNN is a stack of layers defined by the action of kernels on the input. CNN-based denosiers~\cite{zhang2016beyond,mao2016image} have shown impressive results on image denoising.
The basic idea is based on the inference capability of deep architecture and the power of exploring priors underling images via kernel convolution. 
Even GoogLeNet~\cite{szegedy2015going} has shown the merits of employing kernel combination for building classification models, the study of improving denoising performance via optimizing kernel regulation and combination is not deep enough. To build an ideal CNN-based denoiser by relaxing the two terms from CNN internal is still lacking.

In this paper, we aim to build a CNN-based denoiser to learn to relax the two terms naturally instead. Specifically, we embed a new   kernel regulation strategy into network structure to boost the capacity of modeling priors underling data observation, in the mean time, allow that estimating the latent clean image $\mathrm{ \widehat{x} }$ is not heavy dependency on the inference of deep architecture. Moreover, we take a new perspective of transfer learning into account as a strategy to enhance the model generality.
In turn, this is able to relax likelihood term from entire model simultaneously. 
We make the following contributions: 
\begin{itemize}
  \item We propose a kernel regulation module (neural network-block), referred as KR-block, that can effectively estimate features in solving image denoising by placing a small kernel behind a large one and blend the features captured by the small one with the feature-map processed previously through the large kernel. 
  \item We show that the idea of decoupling likelihood terms and priors can be introduced into CNN internal designing via regarding each KR-block as one individual denoiser. Based on this, we build a state-of-the-art deep CNN-based denoiser, referred as KRNET, via concatenating multiple KR-blocks. 

  \item Our proposed KRNET is capable of handling additive white Gaussian noise (AWGN), multi-channel(MC) noise, and realistic noise. Our model obtains significant performance gain over the state-of-the-art methods across wide level noise.
\end{itemize}

\begin{figure*}[t]
\small
  \centering
  \includegraphics[width=\textwidth]{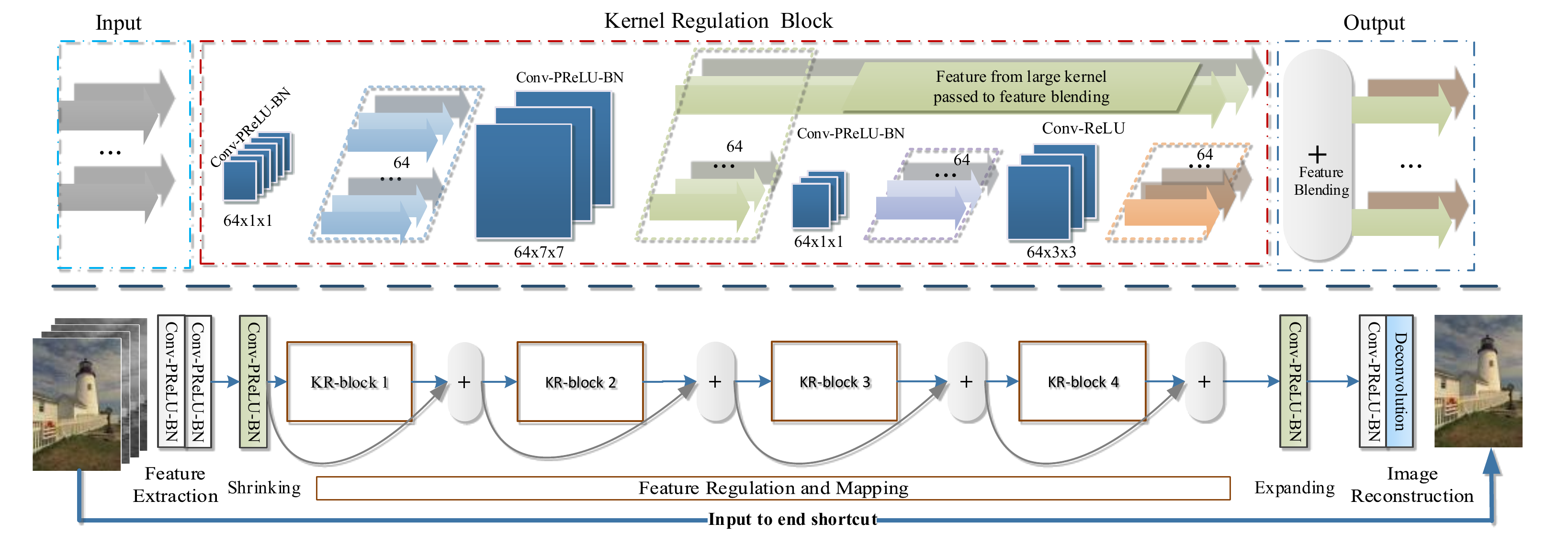}
  \caption{\small Top: A kernel regulation block(KR-block) with a massive of convolution computations (128x7x7) comprises two 1x1convolution components for computation reduction and one 3x3 convolution module for regularizing the features extracted by the preceding large size kernels. The number of dark-green blocks indicates the quantity of kernels in the current convolutional layers, and the size of dark-green blocks represents the size of kernels and the density of convolution.  The color arrows represent the quantity of feature-map outputs. Bottom: KRENET is consisted of feature extraction, shrinking, regulation and mapping, expanding, and image reconstruction.  }
\label{fig:KRNET}
\end{figure*}

\section{Related Work}

In the past decades, many methods have been proposed to overcome the challenges of solving Eqn.~\ref{eq:map4}. 
Typically, they can be classified as either model-based optimization methods or discriminative learning methods. The former methods~\cite{dabov2009bm3d,gu2014weighted,chen2015external} have more flexibility for handling various image restoration tasks but usually employ a high-computing regularization technique, and tend to directly solve Eqn.~\ref{eq:map2}. In contrast, the latter approaches~\cite{chen2017trainable,barbu2009training,schmidt2014shrinkage} usually obtain impressive performance gain but are restricted in generalization capability, and tend to solve Eqn.~\ref{eq:map3} instead with auxiliary parameters and new objective functions that are trained on paired noisy-clean images. From a new perspective, for these two type of methods, 
once all parameters and regularization terms are determined, the form and process of removing noise from $\mathrm{y}$ are essentially equivalent (see Eqn.~\ref{eq:equ}).
 \begin{equation}
 \label{eq:equ}
 \mathrm{\widehat{x}=F(y; \Theta^{*}) }
 \end{equation}
where $\mathrm{F}$ represents transformation or regularization functions that lead $\mathrm{y}$ to $\mathrm{x}$ along with parameters $\Theta^{*}$ that are associated with $\mathrm{F}$ or trade-off thresholds.  

 Based on the image properties being used, existing methods also can be classified as prior based (i.e.,PCLR ~\cite{chen2015external}), sparse coding based~\cite{elad2006image}, low rank based (i.e.,WNNM ~\cite{gu2014weighted}), and deep learning based (i.e., DnCNN ~\cite{zhang2017beyond}
, REDNet ~\cite{mao2016image} ).  Filter-based approaches (e.g.,~\cite{dabov2009bm3d}) methods are classical and fundamental, and many subsequent studies are built on it~\cite{baxes1994digital}. 
Here,  $\mathrm{F}$ may represent the actions of a type of filer (e.g., mean filter) and the techniques serving for model optimization; $\Theta^{*}$ includes the parameters (e.g., filter matrix, trade-off weights) involved in the calculation of $\mathrm{F}$. 

Deep learning based methods~\cite{zhang2017beyond,mao2016image} have shown 
many advantages in learning the function $\mathrm{F}$ mapping $\mathrm{y}$ to $\mathrm{x}$  
by using multi-layer CNNs that are trained on thousands of millions of samples. Here, $\mathrm{F}$ represents the actions of convolution, nonlinear mapping (e.g., ReLU~\cite{krizhevsky2012imagenet}), and  optimization techniques (e.g., Batch Norm~\cite{ioffe2015batch}); $\Theta^{*}$ represents all learned weights in regarding with the actions of $\mathrm{F}$. Nonlinear mapping function allows for a real notion of the depth of CNNs and transforms our raw input features into a space where they are linearly separable. Therefore, nonlinear mapping can be regarded as  a type of optimization technique essentially. It can be seen that two type of methods share common actions in the stage of mapping $\mathrm{y}$ to $\mathrm{\widehat{x}}$.

To incorporate merits of two types of methods is a promising strategy to enhance model performance and generality. 
 Recent work~\cite{zhang2017learning} has employed the half-quadratic-splitting (HQS) algorithm~\cite{geman1995nonlinear} by regarding each single CNN-based denoiser as a prior to enhance generality and performance. Inspired by the another recent work~\cite{xiao2017discriminative}, the discriminative learned models can be transferable as long as it can be used in place of the prior regularizing. We investigate the strategy for relaxing the regularization term and data fidelity term from inside of CNNs.  In particular, We believe that by ``REFEED'' the outputs of middle layers of CNNs into training, it may allow that each component of CNNs can model diversiform priors naturally for enhancing model generality.

\section{Kernel Regulation Network}
We combine a convolution (Conv) layer, a batch normalization (BN) layer, and a Parametric Rectified Linear Unit (PReLU) layer~\cite{He_2015_ICCV} as a \textit{composite unit} in our proposed KR-block,  shown in Figure~\ref{fig:KRNET}-top, which is comprised of four composite units adopting large ($7\times7$), small ($3\times3$), and two $1\times1$ convolution kernels, respectively. The proposed KRNET is built up with multiple KR-blocks.
\subsection{Kernel Regulation Block}

One of the popular network architecture using kernel combination strategy is  GoogLeNet~\cite{szegedy2015going}. Coming up with the proposed inception module, GoogLeNet shows that a creative structuring of layers can lead to improved performance and computationally efficiency. 
Inception module places various sizes of kernels in a parallel form combination. This is able to extract fine gain details in volume, while the larger kernel is able to cover a large receptive field of the input.

In classification tasks, extracted diverse information is able to help with the prediction. However, some points are \textit{varying in image denoising}. 

\textbf{Place in series} Image denoising needs an appropriate spacial transformation between $\mathrm{y}$ and  $\mathrm{x}$. Various information may confuse the  CNN-based denoiser. Image classification needs to summarize diverse information to a linear classifier. On the contrary, image denoising needs to find the most prominent  features for a forward  transformation. Therefore, we adopt three sizes (e.g.,$1\times1$, $3\times3$, and $7\times7$) of kernels in KR-block module. They are placed \textit{in series instead}  to allow the small kernels to regulate the features extracted by the large one.

\textbf{Small behind large} Large kernels (e.g.,$7\times7$)) are able to estimate accurate features by observing a local region with more statistical pixel information. The small kernels (e.g.,$3\times3$) are primarily used for exploiting deeper prior information from the underling feature-maps obtained by large preceding large kernels. The subtle textures are especially highlighted during this regularization procedure. As one is well-known, large kernels are  beneficial to noise removal but smooth the whole image irrespective of its edges or details. Small kernels can reserve subtle textures but inevitably capture noise pixel. Therefore, placing a  \textit{small kernel behind the large one} is a straightforward strategy to enhance the denoiser regularization.

\textbf{Feature blending} 

The features extracted by the large kernel contain both real pixel and noise. The small kernel is able to capture real pixel and ignore much noise, simultaneously. At the end of a KR-block, more real pixel features captured by the small kernel blend with the features extracted by the large kernel such that the real pixel features are highlighted. To allow that the locally highlighted real pixel features share with other neighbor KR-blocks, the feature-blending is processed by pixel-wise summation (see Figure~\ref{fig:KRNET}-top) rather than concatenation (e.g. in GoogLeNet). All in all, this helps with finding the most prominent features for a forward transformation as soon as possible.  Eventually, the output of a KR-block contains \textit{more accurate pixel information with less noise} included. 

\textbf{$\mathbf{1\times1}$ convolution} The special usage of $1\times1$ convolution in KR-block is for two purposes: first, it reduces the dimensions inside KR-block modules, such as the the first $1\times1$ convolution layer; second, it adds more non-linearity by having PReLU immediately after every $1\times1$ convolution and suffers with less over-fitting due to smaller kernel size.
\subsection{KRNET Structure}
Consider a corrupted image ${y_{0}}$ is passed through a convolutional network, which intends to learn a mapping function $\mathrm{F}$ between the corrupted image ${y_{0}}$ and a noise-free image ${x}$. 
The  network contains $\mathrm{L}$ convolution layers (Conv), each of which implements a feature extraction transition:
\begin{equation} \label{eq:1}
{x_{l}=Conv(y_{l}, f_{l},n_{l},c_{l})}
\end{equation}
where $l$ indexes the layer, ${y_{l}}$, $f_{l}$, $n_{l}$, and $c_{l}$ represent the $l$'s input, the filter size, filter number, and channel number, respectively. ${x_{l}}$ are the feature maps extracted from ${y_{l}}$ by $Conv(\cdot )$ , which denotes convolution. 
As the top and bottom layers have different functional attentions ~\cite{zeiler2014visualizing}, the network can be decomposed into three parts (see Figure~\ref{fig:KRNET}-bottom): 
feature extraction; feature regulation and mapping; and image reconstruction.

\textbf{Densely convolutional feature extraction}
We use a considerable amount of large filters in the first two ~\cite{zeiler2014visualizing} convolutional layers 
to extract diverse and representative features for feature mapping and spatial transformation.We define densely convolutional features extracted from the $l^{th}$ layer as:
\begin{equation} \label{eq:2}
x_{l}=Conv(y_{l}, f_{l},n_{l},c_{l})_{f\geq 7\times7,n\geq 128}
\end{equation}
In the proposed KRNET, the first two layers have the same volume: $(f_{l},n_{l},c_{l})=(7,128,1)$ for gray image denoising and  $(f_{l},n_{l},c_{l})=(7,128,3)$ for RGB color image denoising.

\begin{figure*}[t]
   \centering
   \includegraphics[width=\textwidth]{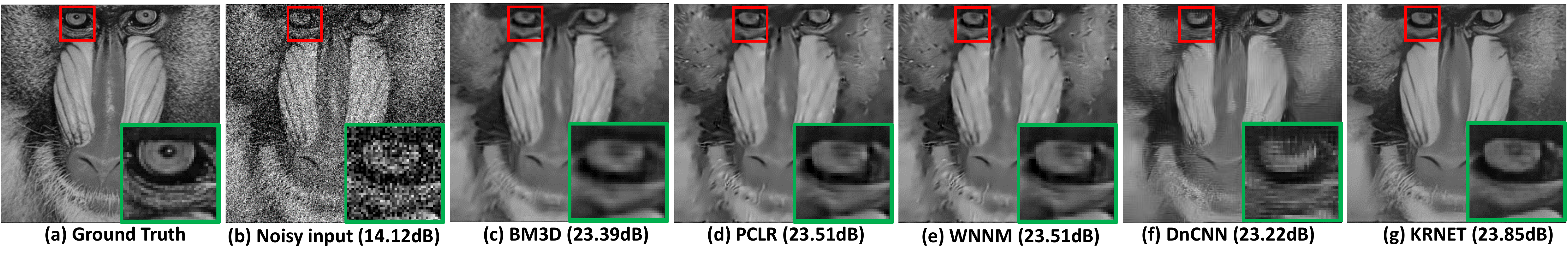}
   \caption{\small Visual results of one image from Set12 test with $\sigma=50$ along with PSNR(dB).}
 \label{fig:N50}
 \end{figure*}

\textbf{Cascade KR-blocks}
Several {KR-blocks} are cascaded to perform feature regulation, mapping, and transformation. Besides, residual learning is performed here by  skip-connection connecting the outputs of two adjacent KR-blocks. The use of skip connection between KR-blocks leads to a faster and more stable training.

\textbf{Input to end shortcut}
The purpose of using a shortcut between the input and the end of the network is to bring more information of the original input to help with the image reconstruction. Because the input data contains much real pixel information that can be taken as a prior for relaxing the network inference difficulty.

\textbf{Scale controlling}
To make KRNET more compact, we introduce two $1\times1$ composite units, referred as ``Shrinking'' and ``Expanding'', shown on Fig~\ref{fig:KRNET}-bottom. After densely convolutional feature-extraction layers, we reduce the number of feature maps by ``Shrinking''.  After feature regulation and mapping, we expand feature maps such that there are sufficient various features that can be provided for image reconstruction.

\textbf{Image reconstruction} One convolutional layer has the volume: $(f_{l},n_{l},c_{l})=(3,128,1)$.  The last layer is a deconvolutional layer with volume: $(f_{l},n_{l},c_{l})=(3,1,1)$  for gray image denoising and $(f_{l},n_{l},c_{l})=(3,1,3)$ for RGB color image denoising.

\begin{table*}[th] \renewcommand\arraystretch{0.2}
\small
  \centering
  \caption{ \label{table:2} {Average PSNR(dB) results of $\sigma$ 10, 30, 50, 70 on BSD200-test dataset~\protect\cite{MartinFTM01}.  The best two results are highlighted in {\color{red}red} and {\color{blue}blue}. Note that KRNET3 has achieved  state of the art.}}
  \begin{tabular}{c|c|c|c|c|c|c|c|c}
  \toprule
  & BM3D & WNNM & PCLR  & DnCNN & REDNET  & KRNET3 & KRNET4 & KRNET5 \\
  & \tiny{\cite{dabov2009bm3d}} & \tiny{\cite{gu2014weighted}} & \tiny{\cite{chen2015external}} &\tiny{\cite{zhang2017beyond}} &\tiny{\cite{mao2016image}} & & &\\
  \midrule
  \small $\sigma = 10 $ & 33.01 & 33.25 & 33.30 &33.25 & 33.63 & 33.97 & {\color{blue}{34.06}} & {\color{red}{34.08}} \\
  \midrule
  $\sigma = 30 $ & 27.31 & 27.48 & 27.54 & 27.50 &27.95 & 28.29 & {\color{blue}{28.48}} & {\color{red}{29.49}} \\
  \midrule
  $\sigma = 50 $ & 25.06 &25.26 & 25.30 & 25.21 & 25.75 & 26.12 & {\color{blue}{26.26}} & {\color{red}{26.29}} \\
  \midrule
  $\sigma = 70 $ & 23.82 &23.95 & 23.94 & 23.93 & 24.37 & 24.77 & {\color{blue}{24.96}} & {\color{red}{25.00}} \\
  \bottomrule
  \end{tabular}
\end{table*}

\begin{table*}  \renewcommand\arraystretch{0.2}
  \centering
  \caption{ \label{table:3}  Average PSNR(dB) results on 24 natural color images of different denoising methods: CBM3D~\protect\cite{dabov2009bm3d}, MLP~\protect\cite{burger2012image}, TNRD~\protect\cite{chen2015learning}, NI~\protect\cite{lebrun2015multiscale}, NC~\protect\cite{lebrun2015noise} and WNNM~\protect\cite{gu2014weighted}. The best two results are highlighted in {\color{red}red} and {\color{blue}blue}.}
  \begin{tabular}{c|c|c|c|c|c|c|c|c|c|c}
  \toprule
  \multicolumn{11}{c}{\small $\sigma_{r}=40,\sigma_{g}=20,\sigma_{b}=30$} \\
  \midrule
   \small CBM3D &\small  MLP &\small TNRD &\small NI &\small NC & \small WNNM-1 &\small WNNM2 &\small  WNNM3 &\small MC-WNNM & \small KRNETMC & \small KRNET-B \\

  \midrule
  \small 27.13 &\small 28.54 &\small 28.68 &\small 25.24 &\small 26.19 &\small 28.84 &\small 28.83 &\small 28.22 &\small29.31 & \small{\color{red}31.85} & \small{\color{blue}30.12} \\
  \midrule
  \multicolumn{11}{c}{\small $\sigma_{r}=30,\sigma_{g}=10,\sigma_{b}=50$} \\
  \midrule
   \small CBM3D &\small MLP &\small TNRD &\small NI &\small NC & \small WNNM-1 &\small WNNM2 &\small WNNM3 &\small MC-WNNM &\small KRNETMC &\small KRNET-B \\
  \midrule
  \small 24.57 &\small 29.27 &\small 29.28 &\small 26.69 &\small 26.61 &\small 29.36 &\small 28.43 &\small 27.52 &\small 30.09 & \small{\color{red}31.85} &\small {\color{blue}30.12} \\
   \bottomrule
   \end{tabular}
\end{table*}

\section{Experiment}
In this section, to prove our model has general ability to recover noisy images
with various types of noise, we perform our model on several denoising tasks: regular
additive white Gaussian noise(AWGN) with zero mean $\mu$ and standard deviation  $\sigma$, multi-channel
additive Gaussian noise which means the $\sigma$ on three channels are different
and real noisy images with noise distribution unknown. 

\subsection{Experiment on regular AWGN removal}

\quad\textbf{Training and Test Datasets.}   Using large images to train model is both time-consuming and not memory friendly since
there is no pooling layer in our network. We follow ~\cite{mao2016image} to crop 300 images from BSD dataset ~\cite{MartinFTM01} into small patches with size of $75\times 75$ for training
under the rule that
the training patch size should be larger than the receptive field. We also explore the effect of
patch size in Section~\ref{sec:ablation}. We train models on four noise levels, $\sigma$ of 10,30,50,70.

Referring to two widely used datasets, we set up test images for performance evaluation of competing methods.
One is the 12 common benchmark images ~\cite{gu2014weighted}. The other is the remaining 200 images from BSD dataset for testing to show generalization of KRNET.

\begin{figure}
\small
  \centering
  \includegraphics[width=0.47\textwidth]{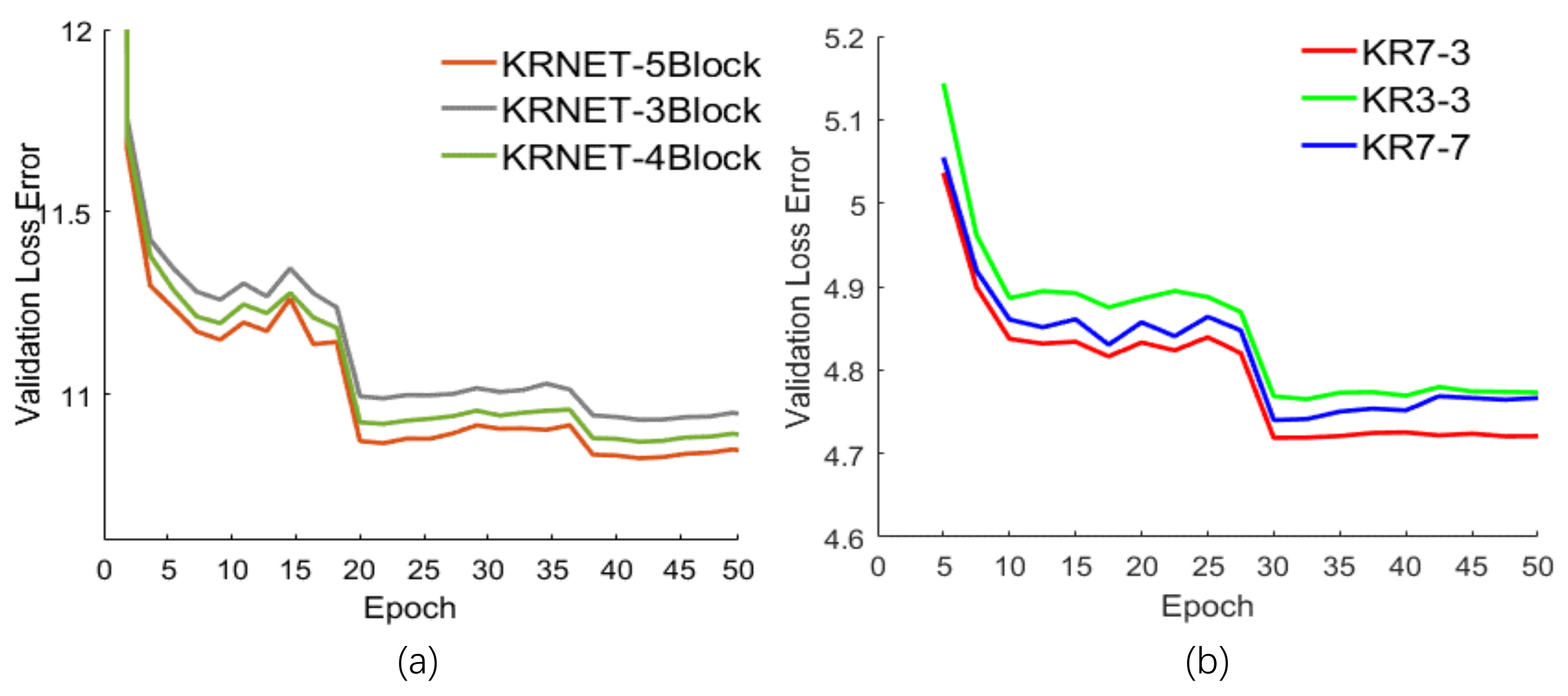}
  \caption {\small(a) Illustration of validation loss. KRNET-5Block, KRNET-4Block, KRNET-3Block denote the models that contain 5,4,3 KR-blocks respectively.
  (b) Illustration of validation loss. KR7-3 denotes the model that contains convolution kernels of multiple size. KR3-3 and KR7-7 denote the variant models that contain kernels of $3 \times 3$ and $7 \times 7$,      respectively.}

  \label{figure:val_loss}
\end{figure}

\begin{figure*}[t]
 \vspace{-1em}
  \centering
  \includegraphics[width=\textwidth]{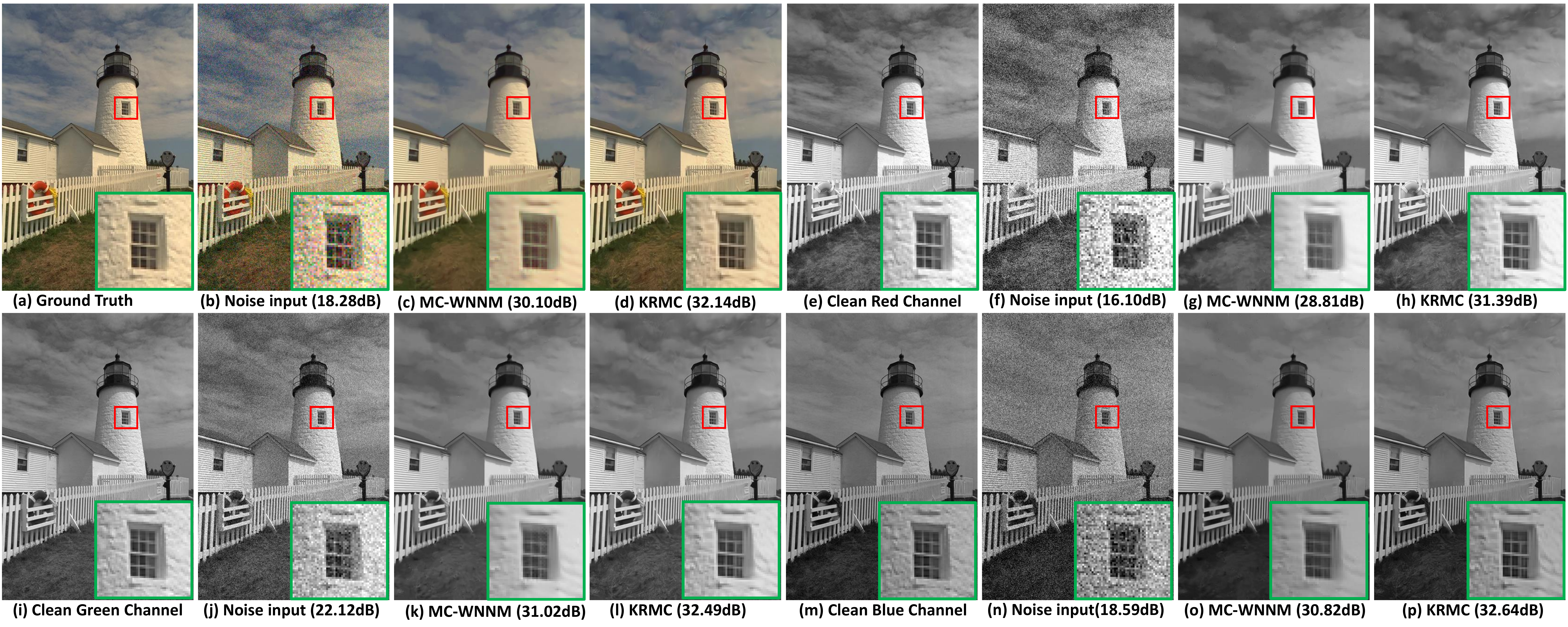}
  \caption{\small Visual results of one image from common set test with multi-channels, (a) to (d) are comparison between MC-WNNM and KRNET on color image, (e) to (h) are same comparison on r channel, as well, (i) to (l) on g channel, and (m) to (p) on b channel}
  \vspace{0.3 em}
\label{fig:MC}
\end{figure*}

\begin{figure*}[t]

  \centering
  \includegraphics[width=\textwidth]{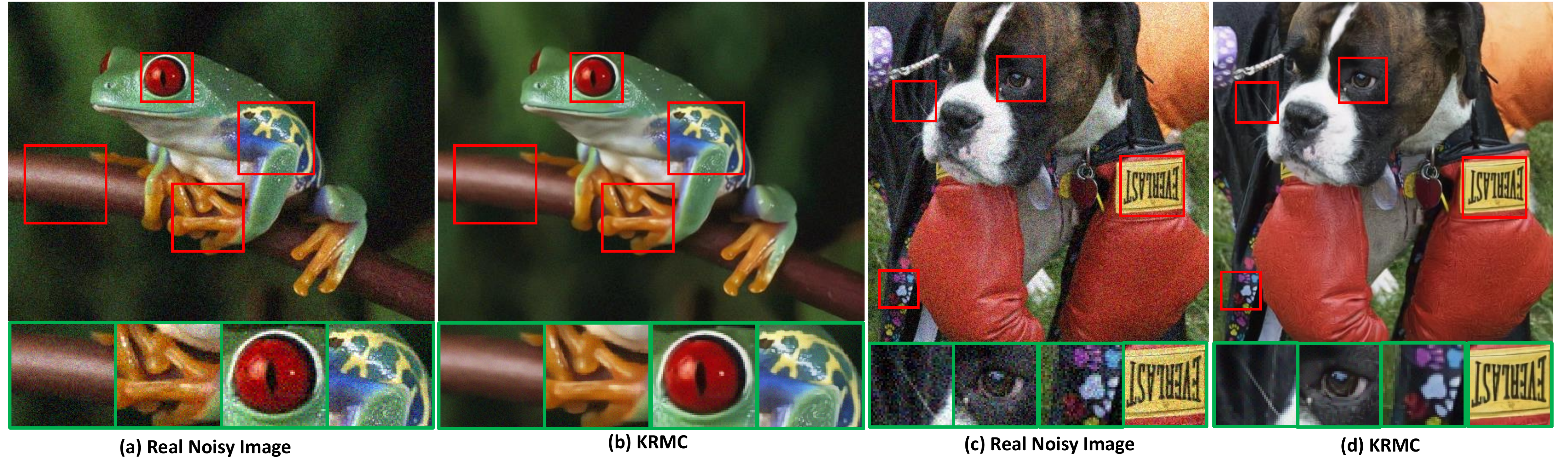}
  \caption{\small Visual results of real noisy image denoising task. Regions of interest are selected where labeled as red squares}

\label{fig:real}
\end{figure*}

\textbf{Implementation details.}
We use the method in ~\cite{he2015delving} to initialize the weights
, $1\mathit{e}-4$ weight decay and 0.9 momentum, and an initial learning rate of $1\mathit{e}-1$, which is decayed exponentially to $1\mathit{e}-4$ for all 50 epochs. 

\textbf{Baseline}  Several state-of-the-art methods are compared with KRNET, including traditional algorithms using different
image properties: filter based method (e.g., BM3D~\cite{dabov2009bm3d}),
 effective prior based method (e.g., PCLR~\cite{chen2015external}), low rank based method (e.g., WNNM ~\cite{gu2014weighted}) and two
 deep learning methods (i.e., DnCNN~\cite{zhang2017beyond}
, REDNET~\cite{mao2016image} ).

\textbf{Quantitative and Qualitative evaluation}  
Table ~\ref{table:2} provides the average PSNR results of different methods on BSD200 dataset.
Compared to the benchmark BM3D, KRNET4(KRNET with 4 KR-blocks) has a notable PSNR gains of about 1.05dB, 1.16dB, 1.21dB, 1.15dB on the four noise levels. 
Moreover, the KRNET4
outperforms the second best methods REDNET by
about 0.5dB for PSNR.
 Qualitative visual comparisons between five different methods have been illustrated in Figure~\ref{fig:N50}, which  is a significantly challenging image due to a great quantity of baboon’s face hair. 
 In this case, our method still outperforms state-of-the-art methods. The KRNET recovered image shows clear contours and sharp edges of eyes, nose, and hair. 


\subsection{Experiment on multi-channel denoising}
\quad\textbf{Experiment details } For real noisy images, the noise statistics in R, G and B channel can be quite different. It is no trivial that MC-WNNM~\cite{xu2017multi} extend the regular AWGN to multi-channel AWGN denoising to fit real image denoising better. 
We crop images from CBSD432 dataset ~\cite{MartinFTM01} into patches of size $65\times 65$ during training. The implementation detail is the same as that on AWGN. We follow~\cite{xu2017multi} to use the same competing
methods and report the PSNR result on  the 24 color images from the Kodak PhotoCD Dataset (http://r0k.
us/graphics/kodak/) by setting $\sigma_{r}=40,\sigma_{g}=20, \sigma_{b}=30 $ and
$\sigma_{r}=30,\sigma_{g}=10, \sigma_{b}=50 $.

In addition to the specific MC-AWGN denoising, we also train a blind MC denoising model, setting the noise level of each channel into the range of $\left [ 0,55 \right ]$ randomly, referred as KRNET-B.

\textbf{Quantitative and Qualitation evaluation}
The comparison results are listed in Table~\ref{table:3}. As one can see, our method achieves 2.54dB, 3.51dB improvements over the state-of-the-art  MC-WNNM and KRNET-B still obtains a gain of 0.69 dB, 2.1dB across two different noise level settings. Visual result is presented in Figure~\ref{fig:MC} as a comparison between our multi-channel KRNET (KRMC) and MC-WNNM. R, G, B channel denoising results are presented separately to indicate the performance on each specific channel.
\subsection{Experiment on real noisy images}
Currently, real noisy images denoising is still challenging due to the lack of ground truth clean images and the irregular distribution of noise.We conduct the KRNET-B model on the images from ~\cite{lebrun2015noise} to assess the
practicability of KRNET. Visual results are illustrated in Figure~\ref{fig:real} . Results show a satisfying performance while removing mixed color noise from the image, the detailed information of textures and contours is completely maintained.

\section{Ablation Experiments} \label{sec:ablation}
\quad\textbf{Structure of KR-block}
To explore the denoising capability of the various kernel combinations, we substitute
 $7\times 7$ kernels with $3\times 3$ kernels in KR-block, denoted as KR3-3 block,
and substitute $3\times3$ kernels with $7\times7$ kernels, denoted as KR7-7 block.
We train these networks with four blocks for $\sigma=30$. Figure~\ref{figure:val_loss} shows the loss curve on validation dataset of the three models during training. It is unsurprising that loss of KR7-7 block, KR3-3 block are both higher than our KR7-3 block. The PSNR of KR7-7 block, KR3-3 block
 on BSD200 dataset is 28.25, 28.32, respectively.

\textbf{Impact of number of KR-block}
To verify the denoising ability of KR-block, we train three models: KRNET3 ( network with 3 KR-blocks ),
 KRNET4 and KRNET5, and report PSNR for AWGN experiment. We also plot the loss curve of three models on $\sigma=70$, shown in Figure~\ref{figure:val_loss}.
From Table~\ref{table:2}, KRNET3 has outperformed the competing methods by a large margin.
Besides, KRNET4 almost has an average gain of about 0.15dB over KRNET3, which demonstrates that the KR-block we propose works well for denoising. However, when network goes deeper,
KRNET5 just outperforms KRNET4 by about 0.03dB, which is reasonable because depth may be not that important for low-level tasks.

 \textbf{Effect of training patch size}
 We train our model on different size of training patches: 45$\times$45, 60$\times$60,
 75$\times$75 on $\sigma=30$, and the corresponding average PSNR on BSD200 dataset is 28.36, 28.41, 28.48, respectively. We conclude that larger patch size is better for the image denoising. The larger size of training patch containing more pixels is better for capturing the latent pixel distribution that is learned by KR-block.

\section{Conclusion}
In this paper, we propose a CNN-based image denoiser, referred to as KRNET, which embeds with KR-block that is designed specifically for image denoising. KR-block incorporates all merits from both large and small kernels and is a novel and concise regularization technique that can enhance the power of modeling prior to internally. 
We show the impressive results of KRNET on Gaussian noise, multi-channel(MC) noise, and realistic noise, respectively. Especially, when tackling with RGB multi-channel color image denoising, KRNET obtains significant performance gain over the existing methods. This is due to the proposed novel feature learning strategy that can relax prior and data fidelity terms to facilitate the network to learn the cross-channel features. The key findings of this work
are two-fold. First,  a well-trained CNN-based denoiser can be regarded as a sequence of filter-based denoisers that are stacked together with the same optimization objective; Second, each component of a CNN-based denoiser have the capacity of modeling diversiform priors naturally.

\bibliographystyle{named}
\bibliography{ref}
\end{document}